\documentclass[11pt]{article}

\usepackage{a4wide}

\usepackage{amsmath,amssymb,amsfonts,color} 

\usepackage{graphics,graphicx}
\usepackage[mathscr]{eucal}

\def\bb{\begin{eqnarray}}
\def\ee{\end{eqnarray}}

\def\tto{{\widehat{\otimes}}}

\newtheorem{lemma}{Lemma}[section]
\newtheorem{satz}{Theorem}[section]

\newcommand{\Proof}{\begin{proof}}
\newcommand{\QED}{\end{proof} \noindent}

\begin{document}
${}$
\begin{center}

{\Large The D-CTC Condition is Generically Fulfilled\\[6pt] in Classical (Non-Quantum) Statistical Systems}
\\[20pt]
{\large {\bf J\"urgen Tolksdorf} and {\bf Rainer Verch}}
\\[16pt]
Institute for Theoretical Physics, University of Leipzig, 04009 Leipzig, Germany
\\[16pt]
rainer.verch@uni-leipzig.de, \ juergen.tolksdorf@uni-leipzig.de
\end{center}
${}$ \\ \\ 
\\ \\
{\small
{\bf Abstract} 
The D-CTC condition, introduced by David Deutsch as a condition to be fulfilled by analogues for processes of quantum systems
in the presence of closed timelike curves, is investigated for classical statistical (non-quantum) bi-partite systems.
It is shown that the D-CTC condition can generically be fulfilled in classical statistical systems, under
very general, model-independent conditions. The central property used is the convexity and completeness
of the state space that allows it to generalize Deutsch's original proof for q-bit systems to more general
classes of statistically described systems. The results demonstrate that the D-CTC condition, or the conditions
under which it can be fulfilled, is not characteristic of, or dependent on, the quantum nature of a bi-partite
system. 
}

\section{Introduction}
\setcounter{equation}{0}
In a seminal paper \cite{Deutsch}, David Deutsch introduced a condition (henceforth referred to as 
{\it D-CTC condition}) that is supposed to capture the meaning of processes ``running back in time''
in bi-partite quantum systems (and more generally, in multi-partite quantum systems, or quantum circuits).
In its simplest form, it can be described as follows: Assume a bi-partite quantum mechnical system given,
consisting of a Hilbert space $\mathcal{H} = \mathcal{H}_A \otimes \mathcal{H}_B$ composed of the Hilbert spaces
of two subsystems. Moreover, suppose that $U$ is a unitary operator on $\mathcal{H}$, which is viewed as 
describing (the result of) a dynamical interaction between the two systems, akin to a time evolution operator, or a scattering matrix. Furthermore,
let $\varrho_A$ be a density matrix on $\mathcal{H}_A$, or synonymously, a state on the ``A''-part of the system with
expectation values 
$$ \langle {\bf a} \rangle_A = {\rm Tr}_A(\varrho_A {\bf a}) $$
for all bounded linear operators ${\bf a}$ on $\mathcal{H}_A$
where we have written ${\rm Tr}_A$ to emphasize that the trace is to be understood with respect to the Hilbert space $\mathcal{H}_A$.
Relative to these data, a density matrix $\varrho$ on
$\mathcal{H}$ is said to {\it fulfill the D-CTC condition} if the following two conditions are fulfilled (see Fig.\ 1 below):
\\[6pt]
(1) \quad The partial state induced by $\varrho$ on the ``A''-part of the system (prior to the interaction $U$ taking
effect) equals
$\langle \,.\,\rangle_A$, 
i.e.\
$$ {\rm Tr}(\varrho ({\bf a} \otimes {\bf 1}_B)) = {\rm Tr}_A(\varrho_A {\bf a}) $$
holds for all bounded linear operators ${\bf a}$ on $\mathcal{H}_A$ (with ${\bf 1}_B$ denoting the identity operator on $\mathcal{H}_B$). 
Equivalently, $\varrho_A$ equals ${\sf tr}^{\mathcal{H}_B}(\varrho)$, the partial trace of $\varrho$ taken with respect to $\mathcal{H}_B$.
Note that in the previous equation, the trace {\rm Tr} appearing on the left hand side is taken on the full Hilbert space 
$\mathcal{H} = \mathcal{H}_A \otimes \mathcal{H}_B$.
\\[6pt]
(2) \quad The partial state induced by $\varrho$ on the ``B''-part of the system (prior to the interaction $U$) returns to 
itself after the interaction $U$ has taken effect, i.e.
$$ {\rm Tr}(\varrho U^*({\bf 1}_A \otimes {\bf b})U)  =  {\rm Tr}(\varrho ({\bf 1}_A \otimes {\bf b})) $$
holds for all bounded linear operators ${\bf b}$ on $\mathcal{H}_B$ 
(with ${\bf 1}_A$ now denoting the identity operator on $\mathcal{H}_A$).
${}$\\
\begin{center}
 \includegraphics[width=8cm]{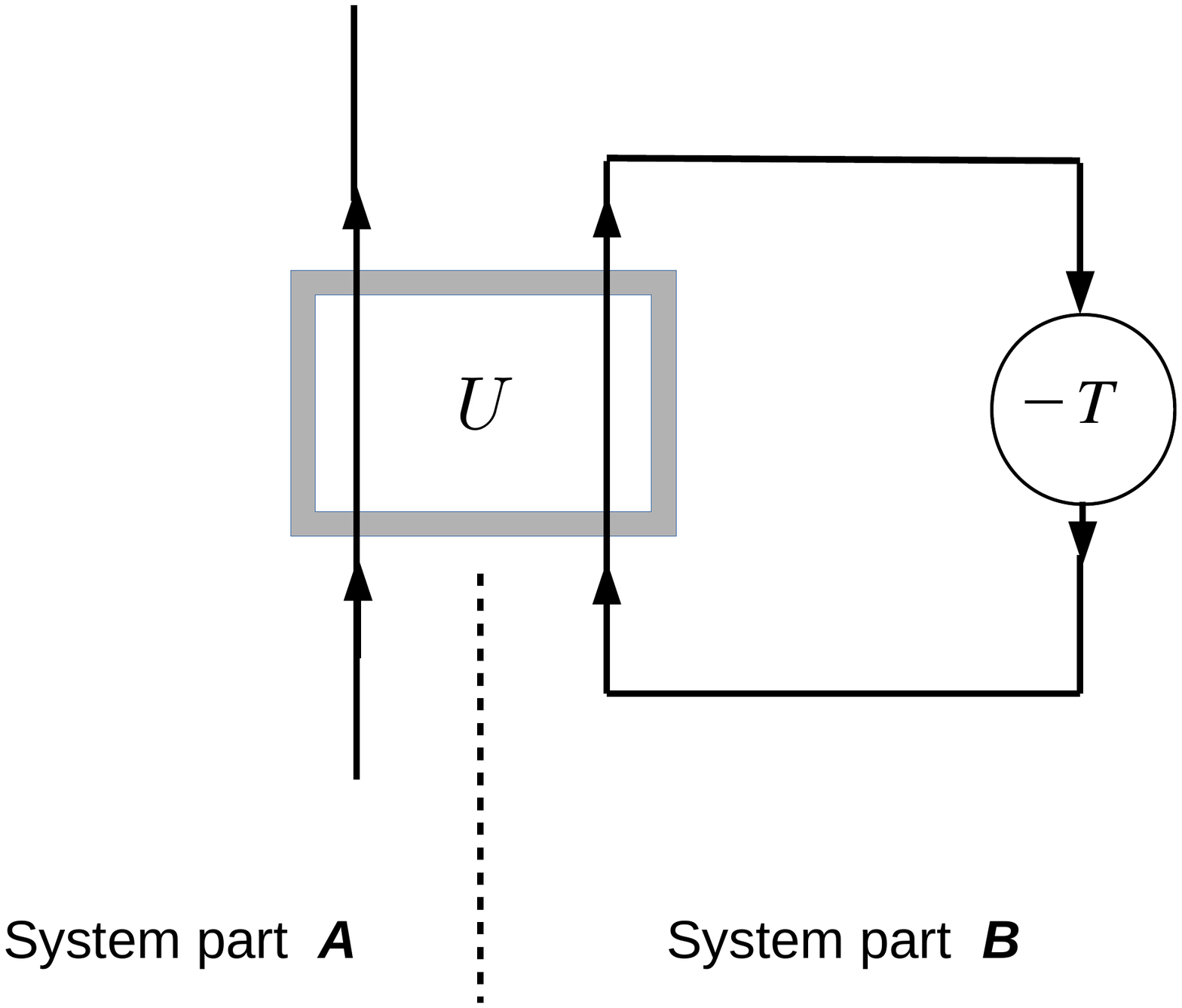}
\end{center}
{\small {\bf Figure 1.} A process in a quantum circuit is represented a unitary operator $U$
describing the dynamical coupling of two system parts (denoted by $A$ and $B$). $U$ takes initial states
(prior to interaction) to final states (after the interaction has taken effect); that process is
supposed to take a time duration $T$.
A ``step backward in time''
is symbolized by $-T$; the $B$-part of the result of the process (i.e.\ the partial state on the 
$B$ system after the interaction) is again fed into the process
as initial state of the $B$-part.}
\\[6pt]
In his proof that the D-CTC condition can always be fulfilled when the Hilbert spaces $\mathcal{H}_A$ and $\mathcal{H}_B$ are both
finite-dimensional, Deutsch uses that the map
$$ \mathcal{S}: \varrho_B \mapsto {\sf tr}^{\mathcal{H}_A}( U(\varrho_A \otimes \varrho_B)U^\ast) $$
on the set of density matrices $\varrho_B$ on $\mathcal{H}_B$
has a fixed point. However, what is actually being used (and allows the fixed point argument to be applied) 
is that for quantum mechanical systems, the state space is always convex and 
complete: It allows for classical statistical (or probabilistic) mixtures of states, and limits (in a suitable sense) thereof. In other words,
the D-CTC condition, and the question to which extent it can be fulfilled, is not primarily sensitive to, or dependent on, genuinely
quantum mechanical properties of a bi-partite system, such as quantum mechanical superpositions (interference effects), uncertainty relations
or entanglement, but really on the convexity and completeness of the state space of the systems in question. Therefore,  {\it the D-CTC condition
can also be fulfilled in classical (i.e.\ non-quantum) statistical physical theories}, such as classical statistical mechanics, under
very general, physically realistic conditions; it is the purpose of this article to demonstrate that fact at an appreciable
level of mathematical generality and rigour.
The authors of \cite{BWW} reach at a related conclusion, however based on a different reasoning
than presented in this article; they argue that in the limit of large Hilbert space dimension of the ``B'' system part, the D-CTC condition
becomes classical.
In \cite{PGC}, the authors indicate that the D-CTC condition can be staged in a far more general formal framework than
that of quantum mechanics.
The feature of the D-CTC condition
to be primarily dependent on the ability to form classical statistical mixtures of states has also been observed in \cite{ArntzeniusMaudlin}.
That same article
also discusses related investigations of classical ``billiard ball''
collisions wherein one of the balls enters a ``wormhole''-type time machine and re-emerges ``prior to entering''
exactly such as to 
be kicked by the other ball into the time machine \cite{FrolovNovikov,E-K-T,Novikov}. Such scenarios may be viewed as particular 
classical counterparts of the D-CTC set-up (or rather -- historically more correctly -- the D-CTC approach ought to be seen as an attempt at providing
a quantum analogue for such ``billiard-ball-collisions-with-wormhole-time-machines'' set-ups)
but we will not follow this line of analogy in the present article. See, however,
Sec.\ 5 for further remarks.

The D-CTC condition is always presented in the context of quantum physics\footnote{The article \cite{DeutschLockwood} 
portrays quantum mechanics as a natural mechanism for avoiding paradoxes that would occur in the presence of CTCs in the 
framework of classical physics; however in \cite{ArntzeniusMaudlin} it is pointed out that forming statistical mixtures would in a similar way 
allow it to avoid those paradoxes. Nevertheless, linking the D-CTC condition with quantum physics has become a commonplace because of 
its origins and potential consequences in quantum computing (see references cited above), and seems to have also gained traction in popular culture: In
\cite{MC:AvEG}, the lead character Tony Stark verbally mentions ``the Deutsch proposition'' in an attempt of the protagonists to 
travel back to the past using a fictional time machine based on quantum physics.}
or of quantum computational considerations (as a sample, see the 
publications \cite{Deutsch,AMRM,BWW,Allen,BrunWilde,RalphMyers,BubStairs,BLSS}, see also references therein).
Therefore, it seems well worth pointing out, and demonstrating, that it is basically of a statistical,
but not necessarily quantum physical nature. 


In a recent paper \cite{jtrv1} (see also \cite{rvDCTC19} for a summary), we have investigated the D-CTC condition in the 
setting of operator-algebraic quantum field theory \cite{Haag}. It is useful to briefly outline some of the basic elements
of that approach as it helps to make parallels between the D-CTC condition as formulated above for quantum mechanical systems, and 
the classical statistical physics case to be considered in this article, more easily visible. 

In the operator-algebraic approach to relativistic quantum field theory \cite{Haag}, there is for any system (quantum field) a $C^*$-algebra
$\mathcal{A}$ whose self-adjoint elements correspond to observables of the system. In most cases, it is no major restriction
to suppose that $\mathcal{A}$ is a subalgebra of some ${\sf B}(\mathcal{H})$, the algebra of all bounded operators on an (infinite
dimensional) Hilbert space $\mathcal{H}$. It is also usually assumed that $\mathcal{A}$ 
contains an algebraic unit element denoted by {\bf 1}; if $\mathcal{A} \subset {\sf B}(\mathcal{H})$, that would be the unit operator on 
$\mathcal{H}$.
Supposing that the spacetime on which the quantum field propagates is Minkowski spacetime
(however, the general setting allows for choosing more general, curved spacetimes instead), it is assumed that for any finite 
(that is, relatively compact)
open region $O$ in spacetime there is a $C^*$ subalgebra $\mathcal{A}(O)$ of $\mathcal{A}$ containing the observables that 
can be measured at times and locations within $O$, including {\bf 1}. In keeping with this set-up, it is further assumed that $\mathcal{A}(O_1) 
\subset \mathcal{A}(O_2)$ whenever $O_1 \subset O_2$. This property is called {\it isotony}. Another assumption is {\it locality},
meaning that ${\bf ab} = {\bf ba}$ for all ${\bf a} \in \mathcal{A}(O_A)$ and ${\bf b} \in \mathcal{A}(O_B)$ provided that 
the spacetime regions $O_A$ and $O_B$ are causally separated, i.e.\ there is no causal curve beginning in $O_A$ and ending
$O_B$. Particularly in this situation where $O_A$ and $O_B$ are causally separated, one may take the pair of algebras $\mathcal{A}(O_A)$
and $\mathcal{A}(O_B)$ as the mathematical model of a causally separated bi-partite system, with $\mathcal{A}(O_A)$ and 
$\mathcal{A}(O_B)$ playing roles analogous to ${\sf B}(\mathcal{H}_A)$ and ${\sf B}(\mathcal{H}_B)$ in the quantum mechanical
setting outlined at the beginning. 

Another important ingredient of the operator-algebraic approach are states. A state is any expectation value functional 
${\bf a} \mapsto \langle {\bf a} \rangle \quad ({\bf a} \in \mathcal{A})$ on the algebra of observables $\mathcal{A}$, and therefore,
by definition, ${\bf a} \mapsto \langle {\bf a} \rangle$ is linear, and fulfills $\langle {\bf a}^*{\bf a} \rangle \ge 0$ for all 
${\bf a} \in \mathcal{A}$, as well as $\langle {\bf 1} \rangle = 1$. Usually, if $\mathcal{A}$ is contained in some ${\sf B}(\mathcal{H})$,
one considers only {\it normal} states which arise from density matrices; in other words, a state is normal if it is of the form
$$ \langle {\bf a} \rangle = \langle {\bf a} \rangle_\varrho = {\rm Tr}(\varrho {\bf a}) \quad ({\bf a} \in \mathcal{A}) $$
for some density matrix $\varrho$ on the Hilbert space $\mathcal{H}$. 

One may now reformulate the D-CTC condition in the operator-algebraic setting as follows. As mentioned, one starts from 
an observable algebra $\mathcal{A} \subset {\sf B}(\mathcal{H})$ for some Hilbert space $\mathcal{H}$, together with 
observable algebras $\mathcal{A}(O_A)$ and $\mathcal{A}(O_B)$ for two causally separated spacetime regions $O_A$ and 
$O_B$, representing the observables of a causally separated bi-partite system. Further data assumed given are a normal
state $\langle {\bf a} \rangle_A = {\rm Tr}(\varrho_A {\bf a}) \quad ({\bf a} \in \mathcal{A}(O_A))$ on $\mathcal{A}(O_A)$
(on the ``A''-part of the full system) induced by a density matrix $\varrho_A$ on $\mathcal{H}$, and a unitary operator $U$ on
$\mathcal{H}$. Given these data, a state $\langle {\bf c} \rangle = {\rm Tr}(\varrho {\bf c}) \quad ({\bf c} \in \mathcal{A})$
is said to fulfill the D-CTC condition if the following two conditions are fulfilled:
\\[6pt]
(I) \quad The partial state of $\langle \,.\,\rangle$ on $\mathcal{A}(O_A)$ coincides with $\langle \,.\,\rangle_A$, i.e.
$$ \langle {\bf a} \rangle = \langle {\bf a} \rangle_A \quad ({\bf a} \in \mathcal{A}(O_A)) $$
(II) \quad The partial state of $\langle \,.\,\rangle$ on $\mathcal{A}(O_B)$ returns to itself after the action of the 
unitary $U$ has taken effect, i.e.\
$$ \langle U^* {\bf b} U \rangle = \langle {\bf b} \rangle \quad ({\bf b} \in \mathcal{A}(O_B)) $$
The analogy with the D-CTC condition with the quantum mechanical case described above should be clear on noting that, since
both $\mathcal{A}(O_A)$ and $\mathcal{A}(O_B)$ are in a defined way subalgebras of the larger $C^*$-algebra $\mathcal{A}$
(or of ${\sf B}(\mathcal{H})$), and all the algebras share the common algebraic unit element ${\bf 1}$,
the ${\bf a} \in \mathcal{A}(O_A)$ here is analogous to the ${\bf a} \otimes {\bf 1}$ above,
and similarly the ${\bf b} \in \mathcal{A}(O_B)$ here is analogous to the $ {\bf 1} \otimes {\bf b}$ above. We mention however
that in general, in quantum field theory the operator algebra generated by $\mathcal{A}(O_A)$ and $\mathcal{A}(O_B)$ in 
${\sf B}(\mathcal{H})$ need not equal (up to identification) the tensor product $\mathcal{A}(O_A) \otimes \mathcal{A}(O_B)$. A precise 
statement would require introducing von Neumann algebras at this point which we shall not embark on. Nevertheless,
there are criteria as to when such an equality actually does hold, known as {\it split property} or 
{\it statistical independence of states}. We will not further discuss these matters here but refer to 
\cite{Haag,Summers-Indep,FewsterSplit} and references 
cited there for full details.
\\[6pt]
The results obtained in \cite{jtrv1,rvDCTC19} are, roughly, as follows (we give here a mainly qualitative description
and refer to the cited references for full details). States fulfilling the D-CTC condition cannot be found if the states 
are also required to fulfill a Reeh-Schlieder like property \cite{Haag} which implies a strong form of entanglement
\cite{VerchWerner}. On the other hand, if the local algebras of observables fulfill the split property just mentioned,
then one can always find states fulfilling the D-CTC condition approximately to any prescribed precision. 
Since the assumptions are met for a wide range of quantum field theories on globally hyperbolic spacetimes which do not
admit closed timelike curves, the latter result makes it doubtful if the D-CTC condition actually relates to quantum 
processes based on the presence of closed timelike curves in the sense of general relativity. The present work casts doubts
on whether the D-CTC condition has quantum physics at its core. We will address these points in a discussion towards the end in Sec.\ 5.
\\[6pt]
We now turn to describing the content of the present work. In Section 2, we will summarize some basics of commutative
$C^*$-algebras which, in an operator algebraic approach, are used as algebras of observables of classical
(non-quantum) statistical systems. The relation to functions (random variables) on locally compact or compact topological 
Hausdorff spaces and probability measures (states) -- through the Riesz Representation Theorem and the Gelfand-Naimark Theorem -- is also discussed. We have relied on the references \cite{KLaasFQT,F-GB-V,Pedersen,HewittStromberg} for our presentation which,
on one hand, is included to make this work self-contained and to introduce the concepts and notation needed, and on the other hand, to explain some points that need to be taken care of when considering limits of states on certain commutative $C^*$-algebras and the question if they still arise from probability measures. We take up on this topic again in Section 3 
where the concept of classical statistical bi-partite systems is introduced. A criterion ensuring that limits of 
sequences of probability measures exist on $C_b(X)$, the $C^*$-algebra of bounded, continuous functions on a locally compact metric space
$X$, and are again probability measures, is provided by Prohorov's Theorem \cite{Prohorov,BilCPM} and we use it in Thm.\ \ref{Thm02}. In our 
Thm.\ \ref{Thm01} presented before in Sec.\ 3, we prove a very general statement to the effect that the D-CTC condition
for classical statistical bi-partite systems is fulfilled but with states in an abstract $C^*$-algebraic sense which
need not be given by probability measures. As indicated, Thm.\ \ref{Thm02} is more specific in that it establishes that the 
states fulfilling the D-CTC condition are given by probability measures
under certain assumptions.
 A simple example in form of a two-body problem 
interacting by a binding central potential is discussed in Section 4 to illustrate properties of the states fulfilling
the D-CTC condition constructed in Thms.\ \ref{Thm01} and \ref{Thm02}. The example will also serve to 
point out a relation to ergodicity. In the last Section, we collect discussion and 
conclusion, relating our results also to other literature.

\section{Commutative $C^*$-algebras and classical statistical systems}

\subsection{Generalities}

Physical systems that are subject to a statistical description of their measurement values, but are {\it classical} in the 
sense of not being quantum systems, have observable algebras which are commutative (or Abelian). Let us denote a generic 
commutative $C^*$-algebra by ${\sf A}$. Commutativity means that ${\sf fg} = {\sf gf}$ for all ${\sf f,g} \in {\sf A}$ and 
consequently, there are no uncertainty relations among the elements of ${\sf A}$ which would be indicative of a quantum theory.
Likewise, there is no entanglement. Assuming that there is a unit element ${\sf 1}$ contained in ${\sf A}$, a {\it state} on 
${\sf A}$ is defined as a linear functional $\langle \,.\,\rangle: {\sf A} \to \mathbb{C}$, ${\sf f} \mapsto \langle {\sf f} \rangle$ 
fulfilling $\langle {\sf f}^*{\sf f} \rangle \ge 0$ (positivity) and $\langle {\sf 1} \rangle = 1$ (normalization). 
We recall the the well-known fact that any state $w$ on a $C^*$-algebra (commutative or not) is norm-continuous:
$|w({\sf f})| \le ||{\sf f}||$ for all ${\sf f}$ where $||{\sf f}||$ is the $C^*$-algebra norm of ${\sf f}$ \cite{Pedersen}.
It is worth mentioning that $C^*$-algebras are algebras over $\mathbb{C}$ (as field of numbers) but 
that, as in quantum mechanics, only their hermitean elements, fulfilling ${\sf f}^* = {\sf f}$, are considered as observables yielding real-valued expectation values $w({\sf f}) = \langle {\sf f} \rangle$ upon evaluation on states. 

We shall now adopt the mathematical notation and denote a state as $w : {\sf A} \to \mathbb{C}$ so that 
$w({\sf f}) = \langle {\sf f} \rangle$ $({\sf f} \in {\sf A})$, since this notation has some advantages.
It is easy to notice that the set of states on
${\sf A}$, henceforth denoted as $ \mathscr{S} = {\sf A}^*_+$, 
is closed under finite convex combinations, i.e.\ if $w_1,\ldots,w_n$ 
are finitely many states on ${\sf A}$ and $\lambda_1,\ldots,\lambda_n$ are 
non-negative numbers such that $\sum_{k =1}^n \lambda_k = 1$, then the convex sum 
$\sum_{k = 1}^n \lambda_k w_k $ that can be formed from the given states 
is again a state on ${\sf A}$. A state is called {\it pure} if it can be represented in this convex sum form if, and only if, all the $w_k$ coincide;
or equivalently, iff all $\lambda_k = 0$ except for exactly one $\lambda_{k'}$ which therefore must be equal to 1. A state which is not pure is 
called {\it mixed}. 
Furthermore, the set of states is closed with respect to taking weak limits: Suppose that $\{w_\kappa\}_{\kappa \in \mathcal{K}}$ is 
a generalized sequence of states $w_\kappa \in \mathscr{S}$, where $\mathcal{K}$ is an arbitrary directed set. The 
generalized sequence $\{w_\kappa\}_{\kappa \in \mathcal{K}}$ 
is called {\it weakly convergent} (strictly speaking, {\it weak-*-convergent})
if $\lim_\kappa \,w_\kappa({\sf f})$
exists for every ${\sf f} \in {\sf A}$. Then 
$$ w({\sf f}) = \lim_\kappa \,w_\kappa({\sf f}) \quad ({\sf f} \in {\sf A}) $$
is again a state on ${\sf A}$. We mention also that $\mathscr{S}$ is weakly compact
by the Banach-Alaoglu-Theorem \cite{ReedSimon1}, which entails that,
whenever $\{w_\kappa\}_{\kappa \in \mathcal{K}}$ is a generalized sequence in $\mathscr{S}$, then
it admits a weakly convergent generalized subsequence $\{w_{\kappa(\zeta)}\}_{\zeta \in \mathcal{Z}}$ (with 
suitable directed index set $\mathcal{Z}$).

An {\it operation} is any map $\tau : \mathscr{S} \to \mathscr{S}$ which preserves convexity, meaning that 
$$ \tau\left(\sum_{k = 1}^n \lambda_k w_k\right) = \sum_{k = 1}^n \lambda_k \tau(w_k) $$
for all finite convex sums of states. 
Moreover, we will assume operations to be weakly continuous\footnote{Strictly speaking, the continuity property defined here
is {\it weak-*-continuity}.} which is defined as follows: 
 $\tau$ is 
 {\it weakly continuous} if, for all weakly converging generalized sequences $\{w_\kappa\}_{\kappa \in \mathcal{K}}$ of 
states in $\mathscr{S}$, also $\{\tau(w_\kappa)\}_{\kappa \in \mathcal{K}}$ is a weakly converging generalized sequence of states,
with $\lim_\kappa\,\tau(w_\kappa)({\sf f}) = \tau(\lim_\kappa\,w_\kappa)({\sf f})$ for all ${\sf f} \in {\sf A}$.
We will see some examples soon; obviously, if $\alpha: {\sf A} \to {\sf A}$ is a $C^*$-algebra 
morphism which preserves the unit element, then its dual map $\alpha^*(w)({\sf f}) = w(\alpha({\sf f}))$ is an operation.

\subsection{The Gelfand-Naimark Theorem}

The next step is to summarize the content of the Gelfand-Naimark theorem (see Lemma 2 in \cite{GelfandNaimark}) which characterizes commutative 
$C^*$-algebras as sets of number-valued functions and the states as probability measures. To this end, we largely follow the presentations of 
\cite{KLaasFQT}, \cite{Pedersen} and \cite{F-GB-V} which we recommend for further reading. 

Let ${\sf A}$ denote a commutative $C^*$-algebra with unit element ${\sf 1}$. Then the Gelfand-Naimark theorem asserts that there 
is a compact topological Hausdorff space $X$ and a $C^*$-algebra isomorphism $\phi : {\sf A} \to C^0(X)$ which preserves the 
unit. Here, $C^0(X)$ is the vector space of all continuous functions on $X$ taking values in $\mathbb{C}$; endowing it with 
the pointwise product $(fg)(x) = f(x)g(x)$ $(x \in X)$ as an algebra product and complex conjugation as the $*$-operation, and 
taking as $C^*$-norm $||f||_\infty = \sup_{x \in X}\, |f(x)|$, $C^0(X)$ is a commutative $C^*$-algebra. Its unit element clearly 
is the function $1(x) = 1$ $(x \in X)$ taking identically the value $1$. Moreover, for any state $w$ on ${\sf A}$, the induced state
$w^\phi(f) = w(\phi^{-1}(f))$  on $C^0(X)$ is given by a probability measure $\mu_w$ defined on the Borel sets of $X$:
$$ w^\phi(f) = \int_X f(x)\,d\mu_w(x) \quad (f \in C^0(X))\,.$$
A probability measure is normalized so that $\int_X 1\, d\mu_w = 1$. Furthermore, a state $w$ on ${\sf A}$ is pure if and only if 
the measure $\mu_w$ is concentrated at a single point $x_0$ in $X$ (a ``Dirac measure''), that is, $w^\phi(f) = f(x_0)$ for all
$f \in C^0(X)$. Therefore, the {\it Gelfand transform} ${\sf f} \mapsto f$, $f(x) = w_x({\sf f})$, where the $w_x$ ($x \in X$) range
over the set of pure states on ${\sf A}$, provides the concrete realization of the isomorphism $\phi$.
Any homeomorphism $F: X \to X$ gives rise to a $C^*$-isomorphism $A_F: C^0(X) \to C^0(X)$ given by 
$A_F(f) = f \circ F^{-1}$ and one has $A_F(1) = 1$; pulling $A_F$ back by $\phi$ renders a $C^*$-algebra isomorphism 
$\alpha_F$ of ${\sf A}$ given by $\alpha_F = \phi^{-1} \circ A_F \circ \phi$ which preserves the unit element ${\sf 1}$.
Consequently, the dual map $\tau_F = \alpha_F^*$ is an operation on the set of states on ${\sf A}$. On the Borel measures
of $X$, this operation is given as $\mu \mapsto A_F^*(\mu) = \mu \circ F$ which can be seen from the measure-transformation
equation (cf.\ \cite{HewittStromberg}, Thm.\ 12.46)
$$ \int_X f \circ F^{-1} \, d\mu = \int_X f \, d(\mu \circ F) \quad (f \in C^0(X))\,. $$ 
Consequently, for a commutative $C^*$-algebra, operations on the set of states arise from bijective homeomorphisms in the indicated way.
There are also operations typically not arising in this way. A simple example is $\tau: w \mapsto \frac{1}{2}(w_0 + w)$ where $w_0$ is any fixed but 
arbitrary state. Another class of examples concerns operations on a particular set of states. Assume that a commutative $C^*$-algebra with unit element
is given as $C^0(X)$ for a compact Hausdorff space $X$, and select any state $w_0$, i.e.\ a probability measure $\mu_0$ on the 
Borel sets of $X$. Then the Hilbert space of the GNS representation (see Thm.\ 1 in
\cite{GelfandNaimark}, or  Thm.\ 3.3.3\ in \cite{Pedersen} for a more modern version)\footnote{Given any $C^*$-algebra $\mathcal{A}$ (not necessarily commutative) containing a unit element, and a state $\omega$ on $\mathcal{A}$, the {\it GNS representation} is a triple $(\mathcal{H},\pi,\Omega)$ where $\mathcal{H}$ is a Hilbert space, $\pi$ is a unital $*$-representation
of $\mathcal{A}$ by bounded linear operators on $\mathcal{H}$ and $\Omega$ is a unit vector in $\mathcal{H}$ so that $\omega({\bf a}) = \langle \Omega,\pi({\bf a})\Omega\rangle$ holds for all ${\bf a} \in \mathcal{A}$, and $\pi(\mathcal{A})\Omega$ is dense in $\mathcal{H}$. For every state on a unital $C^*$-algebra there is such 
a GNS representation and it is unique up to unitary equivalence.
} is given as $L^2(X,\mu_0)$ where at this point, one should bear in mind 
that the $L^2$ space is formed by equivalence classes of square-integrable functions on $X$ where functions are defined as equivalent iff they deviate on sets of zero $\mu_0$ measure. With respect to the chosen $\mu_0$, one can introduce {\it normal states} 
$w_\varrho(f) = {\rm Tr}(\varrho f)$ $(f \in C^0(X))$ where $\varrho$ is any density matrix on the Hilbert space $L^2(X,\mu_0)$
and $f \in C^0(X)$ acts as multiplication operator on $L^2(X,\mu_0)$. Then any unitary linear operator $U$ on $L^2(X,\mu_0)$ induces 
the operation $\tau_U : w_{\varrho} \mapsto w_{U \varrho U^*}$ on the set of normal states with 
respect to $\mu_0$.\footnote{Such an 
operation is in general only weakly {\it sequentially} continuous.}
As a side note, a formulation of classical (statistical) mechanics in a related $L^2$-space setting appears in \cite{Mackey}; it also
serves as a starting point in the so-called geometric quantization \cite{AbrahamMarsden}.

\subsection{The case of ${\sf A} = C_b(X)$ for $X$ non-compact}

The discussion up to now should have clarified the bijective relation between communtative $C^*$-algebras with unit element and their states, and 
the algebras $C^0(X)$ on compact Hausdorff spaces $X$ and the probability measures on the corresponding Borel sets. The latter mathematical framework
is the starting point of classical statistical theories. Here, $X$ usually contains the (classical) degrees of freedom of a physical system;
in fact, most commonly $X = T^*Q$ is the phase space of a system whose degrees of freedom can move in some smooth manifold $Q$. In this situation,
there arises the difficulty that even if $Q$ happens to be compact (which needn't be the case), $T^*Q$ is not. Therefore, we are confronted with
the circumstance that in many physically relevant cases, $X$ isn't in a natural way compact. This issue is of some concern for us because it has 
some consequences for the convergence of states which we need to consider in order to obtain solutions to the D-CTC problem in the following section.

Therefore, assume now that $X$ is a locally compact Hausdorff topological space, and define $C_0(X)$ as the set of all continuous functions
$f :X \to \mathbb{C}$ that {\it vanish at infinity}, i.e.\ given $f \in C_0(X)$, there is for every $\varepsilon > 0$ some compact set 
$K$ such that $|f(x)| < \varepsilon$ for all $x \in X \backslash K$. Using the same definitions for the algebraic operations as for 
$C^0(X)$, the set $C_0(X)$ is a commutative $C^*$-algebra with $C^*$-norm $||f||_\infty = \sup\,\{|f(x)| : x \in X\}$. If $X$ is compact,
then $C_0(X) = C^0(X)$
(cf.\ \cite{HewittStromberg}, Sec.\ (7.13a)),
but if $X$ is not compact, then $C_0(X)$ is a commutative $C^*$-algebra {\it without} an algebraic unit element.
One can still define probability measures $\mu$ on the Borel sets of $X$ as the Borel measures that have unit weight, $\mu(X) = 1$. This 
is equivalent to requiring that the positive functional $w(f) = \int_X f \,d\mu$ ($f \in C_0(X)$) induced by $\mu$ on $C_0(X)$ has unit norm, that is,
$|| w || = 1$ where $||w|| = \sup\,\{|w(f)| : ||f||_\infty = 1\}$. 

The Gelfand-Naimark theorem which has been stated above 
for the case of a commutative $C^*$-algebra ${\sf A}$ with an algebraic unit element has the following extension to the case 
that ${\sf A}$ doesn't possess an algebraic unit element: There is a locally compact Hausdorff space $X$ and a $C^*$-algebraic isomorphism
$\phi: {\sf A} \to C_0(X)$ which again is given by the Gelfand transform; so any state $w$ on ${\sf A}$ (where the normalization condition,
in absence of the algebraic unit ${\sf 1}$, is replaced by the condition that $w$ has unit norm, $||w|| = 1$) induces a state $w^\phi(f) = 
w(\phi^{-1}(f))$ ($f \in C_0(X)$) on $C_0(X)$ which is given by a probability measure on the Borel sets of $X$
(this is exactly the statement of the Riesz' Representation Theorem, see e.g.\ \cite{HewittStromberg}, Thm.\ 12.36), and 
the pure states on ${\sf A}$ are exactly those which arise as probability measures concentrated at single points of $X$.

For any locally compact Hausdorff space $X$, $C_0(X)$ is naturally a $C^*$-subalgebra of $C_b(X)$, the set of all bounded continuous functions 
$f: X \to \mathbb{C}$. Clearly, $C_b(X)$ becomes a $C^*$-algebra using the analogous algebraic operations as defined previously for
$C_0(X)$, and again, $||f||_\infty = \sup\,\{|f(x)| : x \in X\}$ as $C^*$-norm. However, $C_b(X)$ contains an algebraic unit element given by 
the function taking the constant value $1$, similarly as for $C^0(X)$ for a compact $X$. While in the case that $X$ is not compact,
$C_0(X)$ is a proper $C^*$-subalgebra of $C_b(X)$, any state $w$ on $C_0(X)$, by being induced by a probability measure on the Borel sets of $X$,
extends uniquely to a state on $C_b(X)$, complying with the normalization condition $w(1) = 1$.  Since $C_b(X)$ is a commutative $C^*$-algebra 
with an algebraic unit element, by the Gelfand-Naimark theorem it is isomorphic to $C^0(\hat{X})$ for a particular compact Hausdorff space $\hat{X}$, the 
 {\it Stone-$\check{C}$\!ech compactification} of $X$. In fact, for any locally compact Hausdorff space $X$, ``extending'' 
$C_0(X)$ to $C_b(X)$ can be viewed as the ``standard model'' of the Stone-$\check{\rm C}$ech compactification. 
We refer to the references \cite{KLaasFQT,F-GB-V,Pedersen} for further discussion
and references on this point. 

We therefore choose the commutative $C^*$-algebra $C_b(X)$ with a locally compact (but not necessarily compact)
Hausdorff space $X$ as the most suitable and versatile version of an observable algebra for a classical statistical system since.

There are the following rationales for that choice:\\[4pt] 
(i) As already mentioned, we view the phase space $T^*Q$ of a mechanical system as the standard example for $X$, and 
$T^*Q$ isn't compact in a natural way. \\[4pt] 
(ii) $C_b(X)$ contains an algebraic unit element while $C_0(X)$ does not for non-compact $X$. However, having a unit is important since 
it allows to approximate unbounded functions, which often represent important observables such as the Hamilton function $H$ (assumed continuous), by 
elements of $C_b(X)$ at the level of expectation values. Namely, in the presence of a unit $1$, one can form 
the resolvents $(1 + \epsilon H^2)^{-1}$ $(\epsilon > 0)$, and then the functions $H_\epsilon = (1 + \epsilon H^2)^{-1} H$  
are in $C_b(X)$ and for sufficiently regular states one obtains $\lim_{\epsilon \to 0}\,w(H_\epsilon) = w(H)$. Therefore, we see 
$C_b(X)$, possessing a unit element, as preferred since it allows the approximation of unbounded observables in a canonical way.
\\[4pt]
(iii) We wish to explore the D-CTC condition in the setting of classical probability theory where, by definition, the states are 
given by probability measures. As mentioned, any state on $C_0(X)$ is actually induced by a probability measure according to Riesz' theorem,
and it extends to a state, induced by the same probability measure, on $C_b(X)$. Therefore, we are not missing any states 
by choosing $C_b(X)$ as observable algebra instead of $C_0(X)$.
\\[6pt]
It should be obvious that 
the operations on states considered previously for $C^0(X)$ with a compact $X$, in particular those induced by bijective homeomorphisms of $X$, have their
completely analogous counterparts also in the case of $C_b(X)$ with locally compact $X$.

However, if $X$ is not compact, then there are states on the $C^*$-algebra $C_b(X)$ which are not given 
by probability measures on the Borel sets on $X$.
 Consider as a particular example the case $X = \mathbb{R}$, and the 
sequence of states on $C_b(\mathbb{R})$ given by $w_n(f) = f(n)$ ($n \in \mathbb{N}$), i.e.\ the point-measures concentrated at the 
integers. If this sequence of states is restricted to $C_0(\mathbb{R})$, it converges for $n \to \infty$ to the zero-functional,
$\lim_{n\to\infty} w_n(f) = 0$ for all $f \in C_0(\mathbb{R})$. Clearly, this functional is not induced by a probability measure and 
therefore the $w_n$ (or any generalized subsequence) do not converge weakly to a {\it state} on $C_0(\mathbb{R})$. On the other hand, 
by the Banach-Alaoglu Theorem mentioned before, there is a generalized subsequence $\{n(\kappa)\}_{\kappa \in \mathcal{K}}$ 
in $\mathbb{N}$ with
$\lim_\kappa n(\kappa) = \infty$ so that the generalized subsequence  $\{w_{n(\kappa)}\}_{\kappa \in \mathcal{K}}$
of states on $C_b(\mathbb{R})$ converges weakly to a state $w(f) = \lim_\kappa w_{n(\kappa)}(f)$. That state $w$ isn't induced
by a probability measure on the Borel sets of $\mathbb{R}$ since $w(1) = 1$ while $w(f) = 0$ for all $f \in C_0(\mathbb{R})$.
One may argue that such states have pathological properties and therefore aren't induced by probability measures and should not be regarded 
as physically realistic states. In order to make the distinction visible in the notation we will, for the commutative $C^*$-algebra
${\sf A} = C_b(X)$, denote the set of all $C^*$-algebraic states by $\mathscr{S}$ as before, and denote the set of states induced by probability measures by 
$\mathscr{S}^{(P)}$. If $X$ isn't compact, then $\mathscr{S}^{(P)}$ is a proper subset of $\mathscr{S}$.

The arguments leading to the results on the generic solvability of the 
D-CTC problem for classical bi-partite statistical systems that we shall derive in the next section
make considerable use of the convergence of (generalized) sequences of states.
Having made the point that we consider $C_b(X)$ with a possibly non-compact $X$ as algebra of observables, 
we would like to specify criteria ensuring that solutions to the D-CTC problem for classical bi-partite statistical systems
are given by states which 
actually are induced by probability measures. As we will see in the next section, the condition of ``tightness''
on sequences of probability measures in combination with Prohorov's Theorem \cite{BilCPM,Prohorov} provide such criteria.

\section{Classical statistical bi-partite systems and the D-CTC condition}

We define a {\it classical statistical bi-partite system} to consist of a direct product $X = X_A \times X_B$ where 
$X_A$ and $X_B$ are locally compact, Hausdorff topological spaces (or, for one of our results below, metric spaces). Then $X$ is also a locally compact
Hausdorff space (res., metric space). We usually think of $X_A$ and $X_B$ as containing the degrees of freedom of two 
system parts labelled ``A'' and ``B'', e.g.\ $X_A = T^*Q_A$ and $X_B = T^*Q_B$ might be phase spaces over finite dimensional
configuration manifolds of many particle systems. The system parts are independent but can be coupled dynamically.
Then we take as observable algebras of the subsystems ${\sf A}_A = C_b(X_A)$ and ${\sf A}_B = C_b(X_B)$; the 
observable algebra of the full system will be ${\sf A} = C_b(X)$. 
Here, one can think of functions on phase space as the classical example. 

Then ${\sf A}= C_b(X)$ contains the $C^*$-subalgebra ${\sf A}_{\tto}$ generated by all elements $f$ of the form 
$$ f = \sum_{j = 1}^N f_j \otimes g_j $$
where $N \in \mathbb{N}$ and $f_j \in C_b(X_A)$ and $g_j \in C_b(X_B)$ where the tensor product is defined by 
$f_j \otimes g_j(x_A,x_B) = f_j(x_A)g_j(x_B)$ for all $x_A \in X_A$, $x_B \in X_B$. We will also write 
${\sf A}_{\tto} = C_b(X_A) \tto C_b(X_B)$. Note that ${\sf A}_{\tto}$ contains the unit element of ${\sf A} = 
C_b(X)$.

If $w_A$ is a state on $C_b(X_A)$ and $w_B$ is a state on $C_b(X_B)$, then one can define the {\it product state}
$w$ on ${\sf A}_{\tto}$ by setting
$$ w(f \otimes g) = w_A(f)w_B(g) \ \ \quad (f \in C_b(X_A)\,, \ g \in C_b(X_B)) $$
and extension by linearity. In the case that ${\sf A}_{\tto}$ is a proper $C^*$-subalgebra of $C_b(X)$, one can still
extend the product state $w$ to a state on  ${\sf A} = C_b(X)$ (\cite{Pedersen}, Prop.\ 3.1.6) which however need not 
be unique. We call any such state {\it a product state extension} of 
 $w_A$ and $w_B$ (to ${\sf A} = C_b(X)$).

If $w_A$ and $w_B$ are states induced by probability measures $\mu_A$ and $\mu_B$ on the Borel sets of $X_A$ and $X_B$ respectively, then
there is a unique product state $w$  induced by a unique probablity measure $\mu = \mu_A \times \mu_B$, the {\it product measure} of $\mu_A$ and $\mu_B$,
on the Borel sets of $X$. The product measure is determined by 
$$ (\mu_A \times \mu_B)(O \times P)\ = \mu_A(O)\mu_B(P) $$
for Borel sets $O$ of $X_A$ and $P$ of $X_B$ (see Sect.\ 21 in \cite{HewittStromberg}).
${}$
We now turn to our results establishing that the D-CTC condition can be fulfilled in classical statistical bi-partite systems
in great generality. In Thm.\ \ref{Thm01} we prove a statement to this end entirely set in the $C^*$-algebraic framework,
where the states aren't necessarily induced by probability measures. Then we present another version in Thm.\ \ref{Thm02}
where the states are induced by probability measures; it is for this result that we make use of Prohorov's Theorem,
summarized below in this section. In the remark following the statement of Thm.\ \ref{Thm01}, we explain how the 
formulation of the D-CTC condition given here connects to (and is, in fact, more general than) that of 
\cite{Deutsch,jtrv1}. 
\begin{satz} \label{Thm01}
 Let $X = X_A \times X_B$ define a classical statistical bi-partite system where $X_A$ and $X_B$ are locally compact, Hausdorff topological spaces
 \\[2pt]
 Let $\tau : \mathscr{S} \to \mathscr{S}$ be an operation and let $w_A \in \mathscr{S}_A$ be a state (in the $C^*$-algebraic sense) on 
 $C_b(X_A)$. Then there is a state $w \in \mathscr{S}$ on $C_b(X)$ (in the $C^*$-algebraic sense) with the properties
 \begin{align} \label{prt1}
  w(f_A \otimes 1)  & = w_A(f_A) \ \  \quad (f_A \in C_b(X_A)) \quad  \text{and} \\
  \tau(w)(1 \otimes f_B) & = w(1 \otimes f_B) \ \ \quad (f_B \in C_b(X_B))\,. \label{prt2}
 \end{align}

\end{satz}
{\bf Remarks} \\[4pt]
(3.1.A) \ \ In line with the terminology in \cite{jtrv1}, we say that a state $w$ with the properties \eqref{prt1} and 
\eqref{prt2} {\it fulfills the D-CTC condition}, or {\it is a solution to the D-CTC problem}, with respect to the 
given $X = X_A \times X_B$, $\tau$ and $w_A$.
\\[4pt]
(3.1.B) \ \ In the Introduction -- where the D-CTC condition of \cite{Deutsch} has been summarized -- and in
\cite{jtrv1}, the operations are always induced by unitary operators on some Hilbert space on which the algebra 
of observables is represented; in other words, they are of the form $\tau (\langle \,. \,\rangle)
= \langle U^* \,.\,U\rangle$
with a unitary operator $U$. The setting here is more general in that this assumption is not being made. In contrast,
another
assumption on operations which enters here is that the operations are assumed to be weakly continuous in the 
sense described above which is a natural assumption in the context of states on $C^*$-algebras. That is not always
a natural assumption when operations are induced by unitary operators where usually weak {\it sequential} continuity 
is a more natural requirement. Thm.\ \ref{Thm02} below actually only requires weak sequential continuity of 
the operation $\tau$. 
\\[4pt]
(3.1.C) \ \ If both $X_A$ and $X_B$ are compact, then the state $w$ is induced by a probability measure according to the 
Gelfand-Naimark theorem. One can deduce this also from Thm.\ \ref{Thm02} below, since the tightness assumptions 
entering in Thm.\ \ref{Thm02} are automatically fulfilled if both $X_A$ and $X_B$ are compact. 
\\[6pt]
{\it Proof of Thm.\ 3.1.} The proof is analogous to the proof given in \cite{jtrv1} in the operator-algebraic quantum field theory context,
which in turn is based on the idea of the proof by Deutsch for the quantum mechanical case in finite-dimensional Hilbert spaces
\cite{Deutsch}.

Choose any state $w^\circ_B$ in $\mathscr{S}_B$ and define the state $\varphi_1$ in $\mathscr{S}$ by choosing a product state 
extension of $w_A$ and $w_B^\circ$,  thence obeying
$$ \varphi_1(f_A \otimes f_B) = w_A(f_A)w_B^\circ(f_B)\,, \quad \ \ f_A \in C_b(X_A), \ \, f_B \in C_b(X_B)\,.$$
Then define a sequence of states $\varphi_n$ ($n \in \mathbb{N}$) in $\mathscr{S}$ inductively choosing 
product state extensions of $w_A$ and the partial state $f_B \mapsto \tau(\varphi_n)(1 \otimes f_B)$, so that
$$ \varphi_{n + 1}(f_A \otimes f_B) = w_A(f_A)\cdot \tau(\varphi_n)(1 \otimes f_B) \quad (n \in \mathbb{N})\,.$$
Note that, as $\tau(\varphi_n)$ is in $\mathscr{S}$, the partial state $f_B \mapsto \tau(\varphi_n)(1 \otimes f_B)$ is in
$\mathscr{S}_B$ which then implies that one may choose a product state extension $\varphi_{n+1}$ from ${\sf A}_{\tto}
= C_b(X_A) \tto C_b(X_B)$ 
to ${\sf A} = C_b(X)$. Without additional conditions however, the product state extensions might be non-unique. 
The definition of the $\varphi_n$ implies for all $n \in \mathbb{N}$ (notwithstanding their potential non-unique extension
to ${\sf A} = C_b(X)$),
\begin{align*}
 \varphi_{n+1}(f_A \otimes 1)&  = w_A(f_A)\cdot \tau(\varphi_n)(1 \otimes 1) = w_A(f_A)\,, \\
 \varphi_1(f_A \otimes 1) &= w_A(f_A)w_B^\circ(1) = w_A(f_A)\,, 
\end{align*}
entailing $\varphi_n(f_A \otimes 1) = w_A(f_A)$ for all $f_A \in C_b(X_A)$. Moreover, we have 
\begin{align} \label{tau-recur}
 \varphi_{n + 1}(1 \otimes f_B) = \tau(\varphi_n)(1 \otimes f_B) \quad \ \ (f_B \in C_b(X_B))
\end{align}
for all $n \in \mathbb{N}$.

Another sequence of states $w_{(N)}$ in $\mathscr{S}$ ($N \in \mathbb{N}$) will then be defined from
the $\varphi_n$ by an averaging procedure:
\begin{align} \label{wN}
w_{(N)}(f) = \frac{1}{N}\left( \sum_{n = 1}^N \varphi_n(f)\right)  \quad \ \ (f \in C_b(X))\,.
\end{align}
It then follows immediately from the properties of the $\varphi_n$ that 
\begin{align} \label{onA}
 w_{(N)}(f_A \otimes 1) = w_A(f_A) \quad \ \ (f_A \in C_b(X_A)\,, \ \, N \in \mathbb{N})\,,
\end{align}
and furthermore,
using \eqref{tau-recur},
\begin{align}\label{tau-inv}
\left| \tau(w_{(N)})(1 \otimes f_B) - w_{(N)}(1 \otimes f_B) \right|
& = \left| \frac{1}{N}\left( \sum_{n = 1}^N \tau(\varphi_n)(1 \otimes f_B)
 - \varphi_n(1 \otimes f_B)\right)\right| \nonumber  \\
 & = \left| \frac{1}{N}\left(\varphi_{N +1}(1 \otimes f_B)^{{}^{}} - \varphi_1(1 \otimes f_B)\right) \right| \nonumber \\
 & \le \frac{2}{N}||f_B||_\infty \quad \ \ (f_B \in C_b(X_B)\,, \ \, N \in \mathbb{N})
\end{align}
Owing to the Banach-Alaoglu theorem \cite{ReedSimon1} that we have already mentioned in the previous section, there is 
a generalized sequence $\{N_\kappa\}_{\kappa \in \mathcal{K}}$, where $\mathcal{K}$ is some directed index set, such 
that $\lim_{\kappa}\,N_\kappa = \infty$ and such that 
$$ \lim_{\kappa} w_{N_\kappa}(f) = w(f) \ \ \quad (f \in C_b(X)) $$
for some state $w \in \mathscr{S}$.
\\
In view of \eqref{onA} and \eqref{tau-inv}, and observing the assumed continuity of $\tau$ which 
asserts that $\lim_{\kappa}\tau(w_{N_\kappa})(f) = \tau(\lim_{\kappa}w_{N_\kappa})(f)$ for all 
$f \in C_b(X)$, one now obtains that $w$ has the properties claimed in the statement above,
\begin{align}
  w(f_A \otimes 1)  & = w_A(f_A) \ \  \quad (f_A \in C_b(X_A)) \quad  \text{and} \\
  \tau(w)(1 \otimes f_B) & = w(1 \otimes f_B) \ \ \quad (f_B \in C_b(X_B))\,.
 \end{align}
This proves the Theorem. \hfill $\Box$
${}$ \\ \\
For the remaining part of this section, we introduce some definitions, following \cite{BilCPM}.
\\[6pt]
Let $\mu$ be a probability measure on the Borel sets of a metric space $X$.
The measure is called {\it tight} if for any given $\varepsilon > 0$ there is a compact subset $K$ of $X$ such that 
$\mu(X \backslash K) < \varepsilon$.
\\[6pt]
Similarly, a sequence $\{\mu_n\}_{n \in \mathbb{N}}$ of probability measures defined on the 
Borel sets of a metric space $X$ is called {\it tight} if for any given $\varepsilon > 0$ there is 
some compact subset $K$ of $X$ such that 
\begin{align}
 \sup_{n \in \mathbb{N}}\,\mu_n(X \backslash K) < \varepsilon \,.
\end{align}
Let $X_A$ and $X_B$ are metric spaces, $X = X_A \times X_B$, and let $\mu$ be a probability measure on the 
Borel sets of $X$. Then one can define the {\it marginals} of $\mu$,
\begin{align}
 \mu^{(A)}(O) = \mu(O \times X_B)\,, \quad \ \ \mu^{(B)}(P) = \mu(X_A \times P)
\end{align}
for Borel sets $O$ of $X_A$ and Borel sets $P$ of $X_B$. Both $\mu^{(A)}$ and $\mu^{(B)}$ are probability
measures.

For later use, we put on record the following statement (see \cite{BilCPM}, Prob.\ 5.9).
\begin{lemma} \label{marginaltight}
 Let $X_A$ and $X_B$ be metric spaces and let $\{\mu_n\}_{n \in \mathbb{N}}$ be a sequence of probability 
 measures on the Borel sets of $X = X_A \times X_B$. 
 \\[2pt]
 Then $\{\mu_n\}_{n \in \mathbb{N}}$ is tight if and only if the sequences of marginals 
 $\{\mu^{(A)}_n\}_{n \in \mathbb{N}}$ and $\{\mu_n^{(B)}\}_{n \in \mathbb{N}}$ are both tight.
\end{lemma}
${}$\\
We shall also make use of the following result.
\\[6pt]
{\bf Prohorov's Theorem} \cite{Prohorov,BilCPM} \\[2pt]
{\it Suppose that $X$ is a locally compact {metric space} and that $\{\mu_n\}_{n\in\mathbb{N}}$ is a sequence of 
probability measures on the Borel sets of $X$. 

If $\{\mu_n\}_{n\in\mathbb{N}}$ is tight, then it is {\rm weakly relatively compact}: There are a probability measure 
$\mu$ on the Borel sets of $X$ and a subsequence $\{\mu_{n(k)}\}_{k \in \mathbb{N}}$ which converges weakly on
$C_b(X)$ to $\mu$, i.e.
\begin{align}
 \lim_{k \to \infty} \, \int_X f\,d\mu_{n(k)} = \int_X f\,d\mu \quad (f \in C_b(X))\,.
\end{align} }
${}$\\[6pt]
\begin{satz} \label{Thm02}
  Let $X = X_A \times X_B$ define a classical statistical bi-partite system where $X_A$ and $X_B$ are 
  locally compact metric spaces.
 \\[2pt]
 Let $\tau : \mathscr{S}^{(P)} \to \mathscr{S}^{(P)}$ be an operation
 on the state space of $C_b(X)$ induced by probability measures on the Borel sets of $X$, and let
 $w_A \in \mathscr{S}_A^{(P)}$ be a state on 
 $C_b(X_A)$ which is induced by a probability measure $\mu_A$ on the Borel sets of $X_A$, assumed to be tight.
 \\[4pt]
 Suppose also that there is a state $w_B^\circ \in \mathscr{S}_B^{(P)}$ on $C_b(X_B)$ which is induced by 
 a probability measure $\mu^\circ_B$ on the Borel sets of $X_B$, with the property that the 
 sequence of probability measures 
 $$ \mu_B^{(n)}{(P)} = \tau^n(\mu_A \times \mu_B^\circ)(X_A \times P) \quad \ \ (n \in \mathbb{N}) $$
on the Borel sets $P$ of $X_B$ is tight.
\\[4pt]
Then there is a state $w \in \mathscr{S}^{(P)}$ induced by a probability measure $\mu$ on the Borel sets of 
$X$ such that 
\begin{align}
  w(f_A \otimes 1)  & = w_A(f_A) \ \  \quad (f_A \in C_b(X_A)) \quad  \text{and} \label{w-onA} \\
  \tau(w)(1 \otimes f_B) & = w(1 \otimes f_B) \ \ \quad (f_B \in C_b(X_B))\,. \label{tau-wB}
 \end{align}

\end{satz} 
{\it Proof.} Using $w_B^\circ$, the sequence of states $\varphi_n$ and whence, the sequence of 
states $w_{(N)} \in \mathscr{S}^{(P)}$ ($N \in \mathbb{N}$) is constructed as in the proof of Thm.\ 3.1.
It follows easily from the assumptions that the states $w_{(N)}$ on $C_b(X)$ are indeed induced by 
probability measures, denoted $\mu_{(N)}$. We wish to show that the sequence $\{\mu_{(N)}\}_{N \in \mathbb{N}}$
is tight. According to Lemma \ref{marginaltight}, this follows once it is shown that the sequences of 
marginals $\{\mu_{(N)}^{(A)}\}_{N \in \mathbb{N}}$ and $\{\mu_{(N)}^{(B)}\}_{N \in \mathbb{N}}$
are both tight. Making use of \eqref{onA}, one can see that
\begin{align} \label{mu-onA}
 \mu_{(N)}^{(A)} = \mu_A \quad \ \ (N \in \mathbb{N})
\end{align}
and as $\mu_A$ has been assumed to be tight, tightness follows for $\{\mu_{(N)}^{(A)}\}_{N \in \mathbb{N}}$.
Similarly, \eqref{tau-recur} shows that 
\begin{align} \label{muB-N}
 \mu_{(N)}^{(B)}(P) & = \frac{1}{N}\sum_{n = 1}^N \tau^n(\mu_A \times \mu_B^\circ)(X_A \times P) \\
                    & = \frac{1}{N} \sum_{n=1}^N \mu_B^{(n)}(P)
\end{align}
holds for all Borel sets $P$ of $X_B$. Since the sequence $\{\mu_B^{(n)}\}_{n \in \mathbb{N}}$ is by 
assumption tight, the same can easily be concluded for the averaged sequence $\{\mu_{(N)}^{(B)}\}_{N \in \mathbb{N}}$.
Hence, the sequence of probability measures $\{\mu_{(N)}\}_{N \in \mathbb{N}}$ is tight. It can therefore be 
concluded from Prohorov's Theorem that there is a subsequence  $\{\mu_{(N(k))}\}_{k \in \mathbb{N}}$ which 
converges weakly to a probability measure $\mu$ on the Borel sets of $X$. Then \eqref{w-onA} follows from
\eqref{mu-onA}, and \eqref{tau-wB} is obtained using \eqref{tau-inv} in combination with the weak continuity of 
$\tau$ as in the final part of the proof of Thm.\ 3.1.
${}$ \hfill $\Box$

\section{A simple example --- and ergodicity}
The Ces\'aro-type limit which enters in the construction of the state $w$ fulfilling the 
D-CTC condition in Thms.\ \ref{Thm01} and \ref{Thm02} is very reminiscent of the discrete time-step evolution
averaged limit which is a standard way of obtaining invariant states under a transformation.
From this perspective, the construction of $w$ is related to Birkhoff's ergodic theorem (\cite{Silva}, Thm.\ 5.1.1).
In this section, we elaborate a bit on this relation, considering a very simple example:
the two-body problem with a spherically symmetric central potential coupling two point masses (particles)
in Hamiltonian mechanics. Thus, we have $X_A = X_B = T^*(\mathbb{R}^3) \simeq \mathbb{R}^3 \times \mathbb{R}^3$ as
phase spaces for the particles labelled ``A'' and ``B'', with Hamiltonian function
$$ H_\lambda(q_A,p_A;q_B,p_B) = \frac{1}{2m_A}|p_A|^2 + \frac{1}{2m_B}|p_B|^2 + V(|q_A - q_B|) + \lambda V_{\rm ex}(q_A,q_B)$$
with particle masses $m_A$ and $m_B$, and $V: \mathbb{R}_{> 0} \to \mathbb{R}$ a smooth function. 
$V_{\rm ex}(q_A,q_B)$ is an external potential and $\lambda \ge 0$ is a coupling constant.
The coupling constant is introduced mainly to distinguish two cases: $\lambda = 0$, i.e.\ the 
center-of-mass moves freely, and $\lambda = 1$, where the center-of-mass moves under the influcence of the external potential.
We think of Newtonian-type potentials like $V(r) = -\alpha/r$ and $V_{\rm ex}(q_A,q_B) = -(\beta_A/|q_A| + \beta_B/|q_B|)$
where $\alpha,\beta_A,\beta_B > 0$; however more general (binding, central) potentials are also possible. For the Newtonian-like
potentials, we would exclude configurations with $q_A = 0$, $q_B= 0$ and $q_A = -q_B$.

In the case of $\lambda = 0$,
the trajectories of bound states (excluding head-on collision) 
correspond to closed ellipses
on which the particles travel in their configuration spaces around the center-of-mass as focal point. For more general 
binding potentials $V$, perihelion shifts may occur for bound states, so that the trajectories
of the particles in their configuration spaces are rosettae revolving around the center-of-mass in a common orbital plane. 

For $\lambda = 1$, the trajectories of the bound states are approximately similar; 
however the center-of-mass trajectory is almost an ellipse 
with the center of the potential at $q_A = 0$ and $q_B = 0$ as focal point. This case corresponds to a planet (the ``A''-system) with 
a moon or satellite (the ``B'' system) that are bound in the gravitational field of a very heavy central star which 
under the mutual gravitational interaction remains almost at rest and can therefore be effectively described as an external potential.
(For this to be a good approximation, the stellar mass is to be very much larger than $m_A$ and $m_B$, and also 
 $|q_A|$ and $|q_B|$ are to be very much larger than $|q_A - q_B|$.)

Then let $F_t : X_A \times X_B \to X_A \times X_B$ denote the phase flow map for the two-body system with
the Hamiltonian function $H_\lambda$, taking phase space points 
from some ``initial'' time $t_i$ to some ``final'' time $t_f = t_i + t$. It induces the $C^*$-algebra 
isomorphism
$$ A_t f = f \circ F_t^{-1} \ \ \quad (f \in C_b(X_A \times X_B))$$ 
on the phase space functions, and in turn it induces the operation
$$ (\tau_t w)(f) = w(A_t f) $$
on the state space $\mathscr{S}$ of $C_b(X_A \times X_B)$. Note that one also has $\tau_t : \mathscr{S}^{(P)} \to \mathscr{S}^{(P)}$,
i.e.\ it maps the set of probability measures on the Borel sets of $X = X_A \times X_B$ to itself.

We wish to illustrate the significance of the tightness assumption in Thm.\ \ref{Thm02}.
Let us look at phase space points $x_A = (q_{A,i},p_{A,i})$ and $x_B = (q_{B,i},p_{B,i})$ at ``initial'' time $t_i$.
The points correspond to pure states on $C_b(X_A)$ and $C_b(X_B)$, induced by Dirac-measures $\delta_{x_A}$ and 
$\delta_{x_B}$ concentrated at $x_A$ and $x_B$, respectively. The configuration space points
$q_A$ and $q_B$ correspond to the inertial 
coordinates of the particles. 

Assume first that $\lambda = 0$. Then, given any $x_A$, it is always possible to find an $x_B$ so that the resulting 
particle trajectories form a bound system, but in general, the center-of-mass will then move with a constant (non-zero)
velocity. In this case one cannot expect that the sequence of measures\footnote{Here we slightly abuse notation
and identify states with the measures by which they are induced; instead of $\tau_t^n(\delta_{x_A} \times \delta_{x_B})$
we should write, more correctly, the transformed measures $(\delta_{x_A} \times \delta_{x_B}) \circ F_t^n$.}
$\tau_t^n(\delta_{x_A} \times \delta_{x_B})$
$(n \in \mathbb{N})$ will be tight because (i) of the validity of Liouville's theorem and 
(ii) the support of these measures in the $q$-components
remains within a ball of sufficiently large, fixed radius around
the center of mass at time $n\cdot t$, so it moves to infinity as $n \to \infty$. (One could compensate that by re-defining $F_t$ so as to 
subtract the center-of-mass motion, but that re-definition depends on the choice of $x_A$ and $x_B$.) 

Consider now the case $\lambda =1$. Then there are a non-empty open set $Y$ and a compact set $K$  in $X = X_A \times X_B$
so that $F_t^n(Y) \subset K$ for all $n \in \mathbb{N}$. Therefore, whenever one chooses a point $x_A \in X_A$ that 
is in the $A$-component of $Y$, there is some $x_B$ in $X_B$ with $(x_A,x_B) \in Y$. Consequently, one obtains that 
the sequence of marginal measures $P \mapsto \tau^n(\delta_{x_A} \times \delta_{x_B})(X_A \times P)$ $(n \in \mathbb{N})$
is tight because all of these measures have their support in the $B$-component projection of $K$, which is a compact subset of $X_B$.
Hence, in this setting, we can apply Thm.\ \ref{Thm02}. As already explained, $\tau$ is identified with $\tau_t$, and we 
may choose $w_A$ as being induced by a probability measure $\delta_{x_A}$ for a phase space point $x_A$ that is part of 
a bound state in $Y$ as just explained; there is actually a wide range of choice for such $x_A$. Then we may choose 
$w_B^\circ$ as any $\delta_{x_B}$ so that $(x_A,x_B) \in Y$. As discussed, the assumptions of Thm. \ref{Thm02} are 
fulfilled, and there is a state $w$ given by a probability measure on the Borel sets of $X$ so that the properties 
\eqref{w-onA} and \eqref{tau-wB} are fulfilled. On the other hand, if one chooses $\delta_{x_A}$ for $w_A$ as before, but 
selects as $w_B^\circ$ a $\delta_{x_B}$ so that $(x_A,x_B)$ does not correspond to a bound state, one cannot expect that the 
required tightness assumption is fulfilled, by a reasoning similar to the $\lambda = 0$ case before.

In the case $\lambda = 1$ and $(x_A,x_B) \in Y$, let us try to understand the properties of the state $w$ constructed 
in Thm. \ref{Thm02} with the properties \eqref{w-onA} and \eqref{tau-wB}. To this end, to ease the illustration,
we consider a very much simplified situation. We assume that $m_A$ is very much larger than $m_B$ so that the motion
of the ``A'' particle coincides to very good approximation with the center-of-mass motion. Furthermore, we assume that
$|q_A|$ is extremely large so that, even for a very high number of orbits of the ``B'' particle around the ``A'' particle,
the center-of-mass motion is approximately a free motion. This corresponds to a satellite, or
``piece of rock'' (``B'' particle) orbiting a planet (``A'' particle) which is on a very remote orbit around a star.
We assume that the orbital planes are coincident, and that the orbit of the satellite around the planet is, to good 
approximation, circular. 
Then we consider the measures \eqref{muB-N} constructed in Thm.\ \ref{Thm02} for the 
present situation,
\begin{align} 
 \mu_{(N)}^{(B)}(P) & = \frac{1}{N}\sum_{n = 1}^N \tau^n_t(\delta_{x_A} \times \delta_{x_B})(X_A \times P)\,.                 
\end{align}
and change to inertial coordinates $q_A$, $q_B$ in which (according to our simplifying assumptions) the planet is approximately at rest.
The measures $\mu_{(N)}^{(B)}$ depend, of course, on how $t$ is chosen. For the assumed (approximately) circular orbit, let $T$ denote 
the orbital period. There are several cases that one can consider:
\\[6pt]
(i) \ \ \ $t = k T$ for some $k \in \mathbb{N}$. Then $\tau^n_t(\delta_{x_A} \times \delta_{x_B})(X_A \times P)$
is independent of $n$ for all $n \in \mathbb{N}$ since we have (in our approximation) $F_t(x_A,x_B) = (x_A,x_B)$ in this 
case. Therefore, $\mu_{(N)}^{(B)} = (\delta_{x_A} \times \delta_{x_B})(X_A \times P)$ is also independent of $N$:
Applying $\tau_t$ just reproduces the initial phase space points.
\\[6pt]
(ii) \ \ \ $t = kT/\ell$ for some $k,\ell \in \mathbb{N}$. Then $F_t^\ell(x_A,x_B) = (x_A,x_B)$ and hence
$\tau_t(\sum_{j = 1}^{\ell} (\delta_{x_A} \times \delta_{x_B})(X_A \times P) = 
\sum_{j = 1}^{\ell} \tau^j_t(\delta_{x_A} \times \delta_{x_B})(X_A \times P)$. Therefore, 
$\mu_{(m \cdot \ell)}^{(B)}(P) = \mu_{(\ell)}^{(B)}(P)$ for all $m \in \mathbb{N}$. 
\\[6pt]
Thus, in case (i), there is a state $w$ fulfilling \eqref{w-onA} and \eqref{tau-wB} whose partial state
on the ``B''-part of the system,
at initial time $t_i$, is given just by $\delta_{x_B}$. In case (ii), there is a state 
$w$ fulfilling \eqref{w-onA} and \eqref{tau-wB} with partial state on the ``B''-part given by 
$$ \frac{1}{\ell}\sum_{j=1}^\ell\delta_{x_B(jkT/\ell)} $$
where $(x_A,x_B(t')) = F_{t'}(x_A,x_B)$. This ``statistical mixture of phase space points'' can be represented 
as $\ell$ copies of the $B$-particle, i.e.\ $\ell$ identical satellites, each on the same circular orbit, separated in 
position and momenta by a $1/\ell$ fraction of the orbit, so that this phase space distribution gets mapped to itself
under the phase space map $F_{kT/\ell}^\ell$.
Quite clearly, the cases (i) and (ii) correspond to periodic orbits of the (effective) motion of the satellite 
around the planet.
\\[6pt]
(iii) \ \ \ $t = rT$ with $r$ irrational. In this case, there are no ``periods'' in the sequence $\mu_{(N)}^{(B)}$
($N \in \mathbb{N}$) and thus any state $w$ constructed in the proof of Thm.\ \ref{Thm02} with the properties 
\eqref{w-onA} and \eqref{tau-wB} can be deduced to be induced, on the ``B''-part of the system, by a measure
$P \mapsto \mu(X_A \times P)$ which is supported on a dense set on the circular orbit of the satellite around the planet.
This follows since on the circle, the irrational rotations (1) are known to produce dense orbits under successive applications
by a classic result of Kronecker (\cite{Silva}, Thm.\ 3.2.3)  and they (2) are ergodic with respect to the Lebesgue-measure on the the 
circle (\cite{Silva}, Thm.\ 3.5.7).
\\[6pt]
The three cases are illustrated in Figure 2. For an interactive illustration, see \cite{Kuehn}
\begin{center}
\includegraphics[width=10cm]{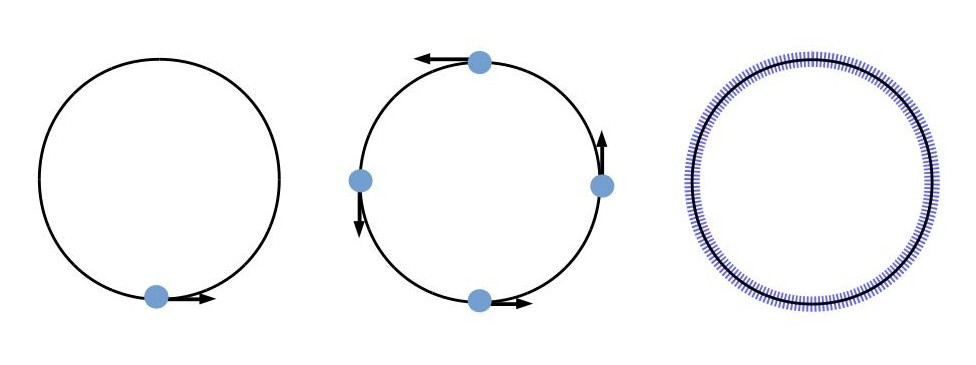}
\end{center}
${}$\\
{\small
{\bf Figure 2.} Illustration of the three cases (i), (ii) and (iii) for the distribution of the state $w$ on the 
``B''-part of the system at initial time $t_i$ mentioned in the text, from left to right. For case (i), the initial
position $q_{B,i}$ of the ``satellite'' on the circular orbit is depicted by the blue dot and its initial momentum $p_{B,i}$
is represented by the arrow. Case (ii) shows a distribution for $k= 4$, with four copies of the ``satellite''
separated by moving the phase space points by $T/4$ on the orbit. In case (iii), one obtains a dense distribution
over the orbit; the momenta are not indicated.}
\\[6pt]
Thus, under the very idealized assumptions made for the simplified situation as described, there is actually a 
state $w$ fulfilling the D-CTC condition for all $F_t$ ($t \in \mathbb{R}$): The ergodic state, obtained by the 
construction of $w$ in Thm.\ \ref{Thm01} for the case that $\tau = \tau_{t'}$ induced by $F_{t'}$ for any 
$t' = rT$ with $r$ irrational. This amounts to taking the Lebesgue measure on the circular trajectory of the satellite.
It is worth pointing out that in 
\cite{PGC} (Appendix B), solutions to the D-CTC problem in a more general formal framework are also obtained by means of an ergodic averaging.

Going back to the days of the very inception of ideas on ergodicity, when thinking of a satellite orbiting a planet,
the rings of the planet Saturn are an example that one might envisage as an approximate realization of 
the ergodic state. (This ``example'' appears in publications of L.\ Boltzmann, see \cite{Gallavotti} for references and 
discussion.) Indeed, if one evolves the ring system by an arbitrary time-step, it appears unchanged, at least at scales larger 
than about 10 km which is tiny compared to the dimensions of the ring orbits -- at scales larger than around 10 km,
the rings, which are mostly formed by rocks and pieces of ice of various sizes between the millimeter and kilometer scale,
appear almost homogeneous in the angular direction (while there are significant density variations in the radial direction)
\cite{Esposito}. (It should be noted that the dynamics of the rings of Saturn is only approximately ergodic, see e.g.\ \cite{Esposito,Bod}
and literature cited there for investigations on this problem.) However, our discussion in this section should serve
to illustrate that the D-CTC condition is nothing extraordinary in classical statistical mechanics, that it relates 
to ergodicity and can be viewed as approximately realized in macroscopic physical systems at appropriate scales.

\section{Discussion and conclusion}

We have shown that the D-CTC condition is generically fulfilled for classical statistical bi-partite systems, under
very general (yet mathematically precise) conditions. The D-CTC condition originated from Deutsch's proposal for giving
a description of what it means that parts of quantum systems undergo processes that might be viewed as analogues to
``going backwards in time''. However, whether or not that condition can be fulfilled rests mainly on convexity and completeness
of the space of states of a system, irrespective of its quantum or classical nature. 
Moreover, in  the light of the results of \cite{jtrv1}, as we 
have indicated in the Introduction, one might also doubt if the D-CTC condition actually says very much about 
closed timelike curves as they are understood in general relativity. This may call into question if the D-CTC 
condition really is a means of ``treating time travel quantum mechanically'' (title of the article \cite{Allen}) or
if statements like ``quantum mechanics therefore allows for causality violation without paradoxes whilst remaining consistent with
relativity'' \cite{RingbauerEtAl} are well-founded. 

The starting point of Deutsch's discussion was
a classical system where the states consist of finite sequences of ``bits'', i.e.\ the state space is a 
discrete, finite set, not admitting convex combinations. That assumption restricts the choice of states fulfillig
the D-CTC considerably, as is shown in an example in \cite{Deutsch}. In contrast, taking ``q-bits'' as the 
``quantized'' version of a classical ``bit'' system naturally renders a convex and complete state space so 
that the D-CTC problem generically has many solutions. Nevertheless, it is rather the possibility to take 
classical statistical mixtures of states than anything specifically quantum mechanical that warrants solutions
to the D-CTC condition in a q-bit system. Allowing classical statistical mixtures of ``bit'' states would have 
the same effect to this end. (However, in applications, ``bit'' states are introduced exactly for the purpose
of avoiding uncertainties in state discrimination that may occur e.g.\ in the form of classical mixtures of states, so that 
from that perspective, forming statistical mixtures doesn't appear natural for 
``bit'' state systems. Yet it is a viable theoretical possibility.)

Therefore, one should be careful not to jump to explanations as to why the D-CTC condition is fulfilled in 
quantum systems which rely on typical features of quantum systems or their behaviour in spacetime (e.g.\ 
interference or uncertainty relations) as this does not relate to what the D-CTC condition -- or the fact that 
it can be generically fulfilled -- is based on; such explanations may result in inadequate interpretations
and are therefore misleading. We would regard the attempt in \cite{Deutsch} to give an explanation for 
the solvability of the D-CTC problem in quantum mechanics based on a many-worlds interpretation, in this sense, as
unconvincing (it has elsewhere been criticized on other grounds \cite{Dunlap}). 

Yet, the fact that the D-CTC condition can generically be solved in q-bit systems can open interesting aspects for
quantum computing and quantum communication \cite{Deutsch,AMRM,BWW,Allen,BrunWilde,RalphMyers,BubStairs,BLSS}. 
In this context, the central point of investigation is -- using the notation
of the beginning part of the Introduction -- the question what ``output'' states on the ``A'' part of the bi-partite system,
$$ \langle {\bf a} \rangle_{\tilde{A}} = {\rm Tr}_A(\tilde{\varrho}_A {\bf a}) = {\rm Tr}(\varrho U^* ({\bf a} \otimes {\bf 1})U)\,,$$
can be derived from the density matrix $\varrho$ fulfilling the D-CTC condition for a given unitary $U$ on the full system, and 
an ``input'' state $\langle {\bf a} \rangle_{A} = {\rm Tr}_A({\varrho}_A {\bf a})$ on the ``A'' part of the system. In other words,
the investigation is on the map $\phi: \varrho_A \mapsto \tilde{\varrho}_A$ for given $U$. There are some difficulties here. First,
since $\varrho$ is not uniquely determined by $U$ and $\varrho_A$, $\phi$ is not naturally defined as a map on the state space
of the ``A'' part of the bi-partite quantum mechanical system. Secondly, given that a map $\phi$ can be determined by imposing
additional selection criteria, if $\varrho$ is constructed as in \cite{Deutsch}, then $\phi$ fails to be convex, i.e.\ it 
isn't an operation. This is also to be expected in the classical (measure-theoretic) framework which we have considered in the present article
in the sense that in general, the dependence of 
the state $w$ in the proof of Theorems \ref{Thm01} and \ref{Thm02} and of 
the partial state $\tilde{w}_A(f_A) = \tau(w)(f_A \otimes 1)$ is not convex in the given state $w_A$ on the 
``A''-part of the system.

That failure of $\phi$ to be convex in $\varrho_A$ is in the literature usually referred to by saying that (solutions to the)
D-CTC condition are ``non-linear'' in the input state $\varrho_A$, and it has been discussed that this may impede the 
utility of the solvability of the D-CTC problem for the purposes of quantum computing. For considerable further investigation
on this issue, see again the articles just cited, and also references given there. A contention expressed in \cite{BWW} is that
due to the failure of $\phi$ to be convex, the D-CTC condition is incomplete. Basically, that is also our conclusion, however
potentially at a more fundamental level, in the sense that the D-CTC condition doesn't depend on typically quantum mechanical features
of a bi-partite system. When claiming that quantum mechanics is an important ingredient in avoiding the notorious
paradoxes of time travel, but then implementing that formally at the level of the D-CTC condition which is not sensitive to whether
a bi-partite system is of quantum mechanical nature or not, and instead just depends on its basic statistical properties, something seems to be missing.

Concerning the possibility that the D-CTC condition isn't sufficiently complete or specific to really allow statements 
connecting quantum processes and closed timelike curves, we have commented in \cite{jtrv1} that a possible 
approach could be to include spacetime localization into the description, in the spirit of the algebraic
framework of quantum field theory as sketched in the Introduction. Still, one would have to connect locality 
properties of the unitary operator $U$, or of the operation $\tau$ with the locality properties of the observables.
In the present paper, we have not considered locality properties of the observables and that is, in a certain sense,
an omission. Therefore, it would be interesting to see if, and how, our results might extend to the analysis of 
billiard ball collisions in the presence of wormhole-type time machines \cite{E-K-T,FrolovNovikov,Novikov}.

In \cite{rvDCTC19}
we have pointed out that the results in \cite{jtrv1} (as well as in \cite{rvDCTC19}) on whether 
the D-CTC condition can be fulfilled or not depend very much on the assumptions made, and on whether 
one insists that the D-CTC condition is fulfilled exactly, or just approximately (to arbitrary precision).
In fact the question of the adequate mathematical idealization is a common problem when trying to 
explore unchartered territory in physics by theoretical methods. In the context of the question if one may
ascribe physical reality to anything which one might bear the properties of a ``time machine'',
i.e.\ processes which can be interpreted as brought about by the presence of closed timelike curves,
the problem of what constitutes an adequate mathematical idealization 
acquires considerable importance, and we think that one of the inspiring aspects about
the investigation of the D-CTC condition is to highlight that issue, and potentially gain some insight.
${}$
\\[20pt]
{\bf Acknowledgement} \ \ The authors would like to thank Christopher J.\ Fewster for comments on the manuscript.


\begin{thebibliography}{22}
\bibitem{AbrahamMarsden} Abraham, R., Marsden, J.E., Foundations of Mechanics, 2nd Edition, Addison-Wesley, 1978
\bibitem{AMRM} Ahn, D., Myers, C.R., Ralph, T.C., Mann, R.B., {\it Quantum state cloning in the presence of a closed timelike curve}, Phys. Rev. {\bf A88} (2013) 022332 
\bibitem{Allen} Allen, J.M.A., {\it Treating time travel quantum mechanically},
Phys. Rev. {\bf A 90} (2014) 042107 
\bibitem{ArntzeniusMaudlin} Arntzenius, F., Maudlin, T., {\it Time travel and modern physics}, Royal Institute of Philosophy Supplement {\bf 50} (2002) 169-200;
revised online version (dated 2009) at Stanford Encyclopedia of Philosophy, https://plato.stanford.edu/entries/time-travel-phys/
\bibitem{MC:AvEG} {\it Avengers: Endgame}, Cinema movie (the citation refers to a scene between 33:00 to 35:00 minutes into the running time),
Marvel Studios (2019) (see https://en.wikipedia.org/wiki/Avengers:${}_{-}$Endgame)
\bibitem{BLSS} Bennett, C., Leung, D., Smith, G., Smolin, J., {\it Can closed timelike curves or non-linear quantum mechanics improve 
quantum state discrimination or help solve hard problems?} Phys. Rev. Lett {\bf 103} (2009) 170502
\bibitem{BilCPM} Billingsley, P., Convergence of Probability Measures, 2nd Ed., John Wiley \& Sons, 1999
\bibitem{Bod} Bodrova, A., Chechkin, A.V., Cherstvy, A., Metzler, R.,
{\it Quantifying non-ergodic dynamics of force-free granular gases}, Phys. Chem. Chem. Phys. {\bf 17} (2015) 21791
\bibitem{BrunWilde} Brun, T.A., Wilde, M.M., {\it Simulations of closed timelike curves},
Found. Phys. {\bf 47} (2017) 375-391
\bibitem{BWW} Brun, T.A., Wilde, M.M., Winter, A., {\it Quantum state cloning using Deutschian closed timelike curves}
Phys. Rev. Lett. {\bf 111} (2013) 190401
\bibitem{BubStairs} Bub, J., Stairs, A., {\it Quantum interactions with closed timelike curves and superluminal signalling},
Phys. Rev. {\bf A89} (2014) 022311
\bibitem{Deutsch} Deutsch, D., {\it Quantum mechanics near closed timelike lines}, Phys. Rev. {\bf D44} (1991) 3197
\bibitem{DeutschLockwood} Deutsch, D., Lockwood, M., {\it The quantum physics of time travel}, Scientific American {\bf 270}(3) (1994) 68-74
\bibitem{Dunlap} Dunlap, L., {\it The metaphysics of D-CTCs: On the underlying assumptions of Deutsch's quantum solution to the paradoxes of time travel},
 Stud. Hist. Phil. Mod. Phys. {\bf 56} (2016) 39-47
\bibitem{E-K-T} Echeverria, F., Klinkhammer, G., Thorne, K.S.,
{\it Billiard balls in wormhole spacetimes with closed timelike curves: Classical theory},
Phys. Rev. {\bf D 44} (1991) 1077-1099
\bibitem{Esposito} Esposito, L.W., Planetary Rings --- A Post-Equinox View, 2nd Ed., Cambridge University Press, 2014
\bibitem{FewsterSplit} Fewster, C.J., {\it The split property for locally covariant quantum field theories in curved spacetime},
Lett. Math. Phys. {\bf 105} (2015) 12, 1633-1661
\bibitem{F-GB-V} Figueroa, H., Gracia-Bond\'ia, J.M., V\'arilly, J., Elements of Noncommutative Geometry,
Birkh\"auser Advanced Texts, Springer-Verlag, 2001
\bibitem{FrolovNovikov} Frolov, V.P., Novikov, I.D.,
{\it Physical effects in wormhole and time machine},
Phys. Rev. {\bf D 42} (1990) 1057-1065
\bibitem{Gallavotti} Gallavotti, G., {\it Ergodicity: a historical perspective. Equilibrium and nonequilibrium}, Eur. Phys. J. {\bf H 41} (2016) 181-259
\bibitem{GelfandNaimark} Gelfand, I., Naimark, M., ``On the imbedding of normed rings into the ring of operators
on a Hilbert space'', Rec. Math. [Mat. Sbornik] N.S. {\bf 12} (1943) 197-217 (open access, 
{\footnotesize
http://www.mathnet.ru/php/archive.phtml?wshow=paper$\&$jrnid=sm$\&$paperid=6155$\&$option${}_{-}$lang=eng)}
\bibitem{Haag} Haag, R., Local Quantum Physics, 2nd Ed., Springer-Verlag, Berlin-Heidelberg-New York, 1996
\bibitem{HewittStromberg} E. Hewitt, K. Stromberg, Real and Abstract Analysis, Springer-Verlag, 
1975
\bibitem{Kuehn} Kuehn, C., {\it An introduction to rotation theory}, online tutorial,
http://tutorials.siam.org/dsweb/circlemaps/circle.pdf
\bibitem{KLaasFQT} Landsman, N.P., Foundations of Quantum Theory, FTPH, Vol. 188, Springer-Verlag, 2017 (open access,
https://link.springer.com/book/10.1007/978-3-319-51777-3)
\bibitem{Mackey} Mackey, G.W., The Mathematical Foundations of Quantum Mechanics, W.A.\ Benjamin, New York, 1963
\bibitem{Novikov} Novikov, I.D., {\it Time machine and selfconsistent evolution
in problems with selfinteraction},
Phys. Rev. {\bf D 45} (1992) 1989-1994
\bibitem{Pedersen} Pedersen, G.K., $C^*$-Algebras and their Automorphism Groups, Academic Press, 1979
\bibitem{PGC} Pinzani, N., Gogioso, S., Coecke, B., {\it Categorical semantics for time travel}, 
34th Annual {ACM/IEEE} Symposium on Logic in Computer Science, {LICS}
               2019, Vancouver, BC, Canada, June 24-27, 2019,
{pp 1--20}, {https://doi.org/10.1109/LICS.2019.8785664}
\bibitem{Prohorov} Prohorov, Y.V., {\it Convergence of random processes and limit theorems in probability theory}, 
Theory Probab. Appl. {\bf 1} (1956) 157-214
\bibitem{RalphMyers} Ralph, T.C., Myers, C.R., {\it Information flow of quantum states interacting with closed timelike curves}, Phys. Rev. {\bf A82} (2010) 062330 
\bibitem{ReedSimon1} Reed, M., Simon. B., Methods of Modern Mathematical Physics, Vol.\ 1, Academic Press,
New York, 1975
\bibitem{RingbauerEtAl} Ringbauer, M., Broome, M.A., Myers, C.R., White, A.G., Ralph, T.C., {\it
Experimental simulation of closed timelike curves}, Nature Communications {\bf 5} (2014) 4145 
\bibitem{Silva} Silva, C.E., Invitation to Ergodic Theory, American Mathematical Society (Student Mathematical Library, Vol.\ 42), 2007
\bibitem{Summers-Indep} Summers, S.J., {\it On the independence of local algebras in quantum field theory},
Rev. Math. Phys. {\bf 2} (1990) 201-247
\bibitem{jtrv1} Tolksdorf, J., Verch, R., {\it Quantum physics, fields and closed timelike curves: The D-CTC condition in quantum field theory},
Commun. Math. Phys. {\bf 357} (2018) 319-351
\bibitem{rvDCTC19} Verch, R., {\it  The D-CTC condition in quantum field theory}, in: Progress and Visions in Quantum Theory in View of Gravity, Finster, F., Giulini, D., Kleiner, J., Tolksdorf, J. (Eds.), pp 221-232, Birkh\"auser, Cham, 2020
\bibitem{VerchWerner} Verch, R., Werner, R.F.,
{\it Distillability and positivity of partial transposes in general quantum field systems},
Rev. Math. Phys. {\bf 17} (2005) 545-576

\end{thebibliography}
\end{document}